\documentclass[aip,graphicx,jcp]{revtex4-1}

\usepackage{graphicx}
\usepackage{rotating}
\usepackage{array}
\usepackage{mathtools}
\usepackage{multirow}
\usepackage[sectionbib]{chapterbib}
\usepackage{verbatim}
\usepackage{braket}
\usepackage{dsfont}
\usepackage{[hysics}
\DeclareGraphicsExtensions{.pdf}

\emergencystretch=\maxdimen
\draft 

\newlength{\intwidth}
\DeclareRobustCommand{\fpint}[2]
   {\mathop{%
      \text{%
        \settowidth{\intwidth}{$\int$}%
        \makebox[0pt][l]{\makebox[\intwidth]{$-$}}%
        $\int_{#1}^{#2}$}}}

\begin{document}

\newenvironment{theprefixbibliography}[2]
  {\begin{thebibliography}{#2}
   \def\@biblabel##1{[#1-##1]}%
   \def\@bibitem##1{\item\if@filesw \immediate\write\@auxout
     {\string\bibcite{##1}{#1-\the\value{\@listctr}}}\fi\ignorespaces}}
  {\end{thebibliography}}
\makeatother


\title{Feshbach projection XMCQDPT2 model for metastable electronic states} 



\author{Alexander A. Kunitsa}
\author{Ksenia B. Bravaya}
\email{kbravaya@gmail.com}
\affiliation{Department of Chemistry, Boston University, Boston, Massachusetts, \nolinebreak{02215, USA}}


\begin{abstract}
Autoionizing electronic states are common intermediates in processes initiated by electron impact or high-energy radiation. These states belong to the continuous spectrum of the Hamiltonian, and as such cannot be treated with methods developed for bound electronic states. Here we propose a new model for describing metastable electronic states, which combines discretized Feshbach projection formalism and multireference perturbation theory and exploits absorbing potential to generate the basis of the coupled valence state and continuum. The results of benchmark calculations for a series of shape resonances in polyatomic molecules  are in good agreement with experimental  and theoretical reference values. 

\end{abstract}

 
\maketitle 

\section{Introduction}
Electronic states metastable with respect to electron ejection (electronic resonances) are commonly formed upon core excitation and ionization ~\cite{Ueda:ETMD:11, Cederbaum:ICD:97,Cederbaum:ETMD:01,NICOLAS:core-ex:12}, as valence excited states lying above ionization continuum~\cite{Bochenkova:Ang:14,Verlet:pBQN,west_anion_2014,Kunitsa:bqn_res:15}. They have been also discussed as precursors of stable anions in the interstellar medium and planetary atmospheres~\cite{Gianturco:ISM:13,Baccarelli2013,Fontenberry:15}. Electronic resonances belong to the continuous spectrum of the electronic Hamiltonian, and, therefore, cannot be described with the techniques developed for bound electronic states. 

During the last decade, significant progress has been made in extending conventional bound state electronic structure methods to treatment of resonances using non-Hermitian quantum mechanics techniques, e.g. using exterior complex scaling~\cite{White:cbs:14,White:CBF:EOM} or complex absorbing potential (CAP) approaches~\cite{Cederbaum:csADC:02,Pal:CAPFSCC:05,Sajeev:CAPCIP:05,Sommerfeld:CAPCI:01,Sommerfeld:CAP-SACCI:12,Ernzerhof:CAPDFT:12,Pal:CAPEOMCC:12,Zuev:CAP:14,Jagau:CAP:13,Kunitsa:xmcq:16,Shiozaki:cap-mscaspt2}. CAP provides a practical tool for extending an electronic structure method to treatment of metastable electronic states by augmenting electronic Hamiltonian with a purely imaginary absorbing potential:
\begin{equation}
\label{eq:capH}
H_{CAP}=H-i\eta W
\end{equation}
where $\eta$ and $W$ determine the strength and the functional form of the absorbing potential, respectively~\cite{Meyer:CAP:93}. Resonance appears as a single square-integrable eigenstate of the CAP-augmented non-Hermitian Hamiltonian (Eq.~\ref{eq:capH}) with a complex eigenvalue, $E=E_R-i\Gamma/2$, where $E_R$ and $\Gamma$ yield estimates of the resonance position and width, respectively.  The electronic structure methods combined with CAP technique include, but are not limited to, equation-of-motion coupled-cluster with singles and doubles excitations, EOM-CCSD~\cite{Pal:CAPEOMCC:12,Zuev:CAP:14,Jagau:CAP:13,cs_review:17}, adiabatic diagrammatic construction, ADC~\cite{Cederbaum:csADC:02}, configuration interaction~\cite{Sajeev:CAPCIP:05,Sommerfeld:CAPCI:01} and symmetry-adapted-cluster configuration interaction~\cite{Sommerfeld:CAP-SACCI:12}, multiconfigurational perturbation theory~\cite{Kunitsa:xmcq:16,Shiozaki:cap-mscaspt2}, and density functional theory~\cite{Ernzerhof:CAPDFT:12}. While the approaches show promising results in computing resonance position and width,  there are still remaining challenges. The major shortcoming that hampers the use of the methods by a non-advanced user is its non-black-box nature. Specifically, the results are rather sensitive to the CAP parameters and, moreover, one has to compute and analyze so-called $\eta$-trajectory, the dependence of complex energy on the strength parameter, $\eta$. The resonance position and width are associated with the real and imaginary parts of the complex eigenvalue corresponding to the stationary point on the $\eta$-trajectory~\cite{Meyer:CAP:93}. Different criteria were proposed to locate the stationary point: either using minimum logarithmic velocity criteria, $|\eta \frac{dE}{d\eta}| \rightarrow min$~\cite{Meyer:CAP:93}, or searching for a stationary point of real and imaginary parts of energy independently~\cite{Jagau:CAP:13}. Yet, in a finite one-electron basis $\eta$-trajectories often exhibit multiple stationary points~\cite{Kunitsa:bqn_res:15}. In addition, is it often hard to distinguish between the trajectories associated with the resonance and discretized continuum states~\cite{Kunitsa:bqn_res:15}.  Note that for most of the CAP-based electronic structure methods one has to perform multiple electronic structure calculations (50-100) to evaluate parameters of a single resonance, i.e. complex eigenvalues of CAP-augmented Hamiltonian have to be evaluated at different values of the $\eta$ parameter. Therefore, the techniques that avoid evaluation and analysis of $\eta$-trajectories are desirable. 

Here we present extended multiconfirgurational quasidegenerate perturbation theory of second order, XMCQDPT2~\cite{XMCQDPT2}, for resonances formulated using discretized Feshbach projection formalism. The projected basis (localized state and continuum) is generated using real absorbing potential of functional form identical to that of $W$ commonly used for CAP calculations (Eq.~\ref{eq:capH})~\cite{Meyer:CAP:93,cs_review:17}. The model reliably reproduced the resonance position and width for a set of shape resonances in diatomic and polyatomic molecules.

The structure of the manuscript is as follows. In Sec.~\ref{sec:theory} we outline Feshbach projection formalism (Sec.~\ref{sec:FP}), discuss the relevant aspects of XMCQDPT2 theory (Sec.~\ref{sec:XMCQ}), and finally present the new model combining Feshbach projection technique with XMCQDPT2 method (Sec.~\ref{sec:cap-xmcq}). The performance of the model is discussed in Sec.~\ref{sec:results}, and the main conclusions are summarized in Sec.~\ref{sec:conclusions}.

\section{theory}
\label{sec:theory}
In this section, we discuss the most relevant aspects of discretized Feshbach projection formalism (Sec.~\ref{sec:FP}) and XMCQDPT2  (Sec.~\ref{sec:XMCQ}) theory that pertain to development of combined theory for metastable electronic states, presented in Sec.~\ref{sec:cap-xmcq}.

\subsection{Discretized Feshbach projection formalism}
\label{sec:FP}
Here we outline the discretized Feshbach projection formalism mainly following the discussion in Refs.~\onlinecite{Feshbach:62,Yanez:87:Feshbach}. In Feshbach projection formalism, the wavefunction of the autoionizing state is represented in the basis of localized states and scattering states using corresponding projection operators, $\mathbf{Q}$ and $\mathbf{P}$, respectively:
\begin{equation}
\ket{\Psi}=\mathbf{Q}\ket{\psi}+\mathbf{P}\ket{\psi}=\sum_{n}{\ket{\Phi_{n}}\braket{\Phi_{n}|{\psi}}}+\int{dE\ket{\Phi_E}\braket{\Phi_E|\psi}}
\label{eq:psi}
\end{equation}
Note that the $\{\Phi_n\}$ and $\{\Phi_E\}$ in Eq.~\ref{eq:psi} are $L^2$-normalized ($\braket{\Phi_n | \Phi_m}=\delta_{nm}$)   and energy-normalized ($\int{dE\Phi^*_E\Phi_{E'}}=\delta(E-E')$), respectively. With this definition of the projection operators the time-indenendent Schr\"{o}diner equation ($\hat{H}\ket{\Psi}=E\ket{\Psi}$) can be rewritten as follows: 
\begin{eqnarray}
(H_{PP}-E)\mathbf{P}\ket{\Psi}=-H_{PQ}\mathbf{Q}\ket{\Psi}
\label{eq:hpp}\\
(H_{QQ}-E)\mathbf{Q}\ket{\Psi}=-H_{QP}\mathbf{P}\ket{\Psi}
\label{eq:hqq}
\end{eqnarray}
Solving Eq.~\ref{eq:hqq} formally for  $\mathbf{Q}\ket{\Psi}$ and plugging in the result in Eq.~\ref{eq:hpp} one arrives to the following result:
\begin{equation}
(H_{PP}-E)\mathbf{P}\ket{\Psi}=H_{PQ}\frac{1}{H_{QQ}-E}H_{QP}\mathbf{P}\ket{\Psi}
\label{eq:ppsi}
\end{equation}
Assuming that $\{\ket{\Phi_n}\}$ are eigenstates of $H_{QQ}$ and that $\ket{\Phi_1}$ represents the state of interest, by introducing introducing optical potential
$V_{opt}=-\sum_{n\neq 1}{\frac{H_{PQ}\ket{\Phi_n}{\bra{\Phi_n}H_{QP}}}{E_n-E}}$,
one arrives to the following non-homogeneous equation for the scattering part of the wavefunction:
\begin{equation}
(H_{PP}+V_{opt}-E)\mathbf{P}\ket{\Psi}=\frac{H_{PQ}\ket{\Phi_1}{\bra{\Phi_1}H_{QP}}}{E_1-E}\mathbf{P}\ket{\Psi}
\label{eq:ppsi}
\end{equation}
Defining $H'=H_{PP}+V_{opt}$ one first finds solution of the homogenous equation:

\begin{equation*}
(H'-E)\mathbf{P}\ket{\tilde{\Phi}_E}=0
\end{equation*}
The solution of Eq.~\ref{eq:ppsi} is then found using Greens' function methods. Comparing the results with the Breit-Wigner expression for a resonance amplitude, one arrives to the following expressions of resonance position and width.  

\begin{eqnarray*}
\Gamma = 2\pi |\bra{\Phi_1}H_{QP}\ket{\tilde{\Phi}_E}|^2\\
E_{R}=E_1+\fpint{}{}dE\frac{\bra{\Phi_1}H_{QP}\ket{\tilde{\Phi}_E}\bra{\tilde{\Phi}_E}H_{QP}\ket{\Phi_1}}{E_1-E}
\end{eqnarray*}
where $\fpint{}{}$ stands for the Cauchy's principal value of the integral. Note that the expressions were derived assuming an isolated resonance.

The above expressions for resonance position and width were obtained under the assumption that $\tilde{\Psi}_E$ are scattering states normalized to delta function. In practice, if Feshbach formalism is to be used together with conventional electronic structure calculations, the scattering states are square-integrable owing to Gaussian-type basis sets. Thus, instead of continuum of scattering states one operates with discretized square-integrable states. In this case, the equations above should be modified as follows~\cite{Yanez:87:Feshbach}:

\begin{eqnarray}
\Gamma = 2\pi |\bra{\Phi_1}H_{QP}\ket{\bar{\Phi}_{E_1}}|^2\rho(E_1)
\label{eq:width}\\
E_{R}=E_1+\sum_{k}{\frac{\bra{\Phi_1}H_{QP}\ket{\bar{\Phi}_{E_k}}\bra{\bar{\Phi}_{E_k}}H_{QP}\ket{\Phi_1}}{E_1-E_k}}
\label{eq:energy}
\end{eqnarray}
where $\rho(E_1)$ density of the scattering states at $E=E_1$ and $\{ \bar{\Phi}_E \}$ states are $L^2$-integrable discretized continuum states, $\braket{\bar{\Phi}_{E_k}|\bar{\Phi}_{E_l}}=\delta_{kl}$.

\begin{figure}[!tbh]
\includegraphics[width=5cm]{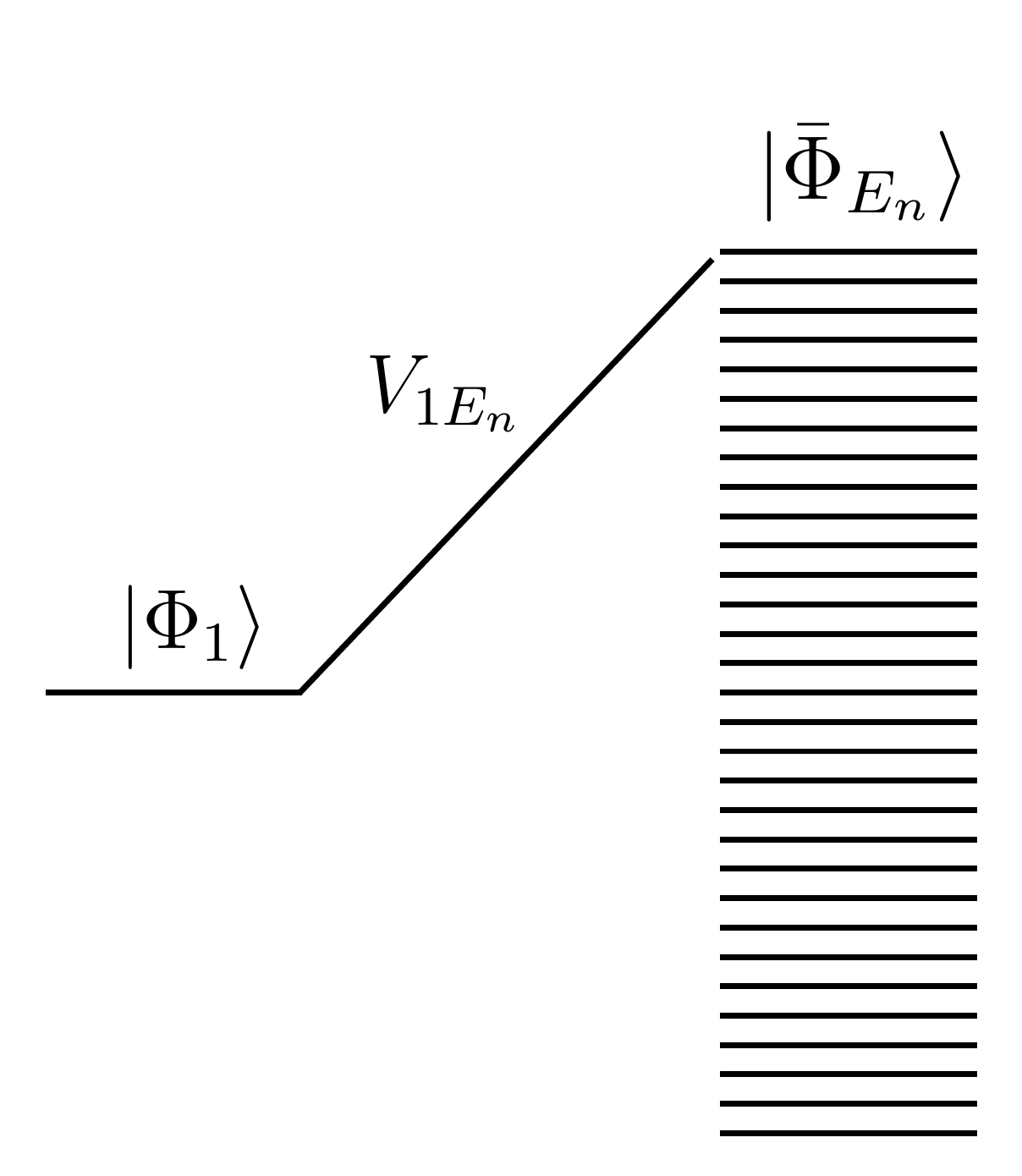}
\caption{\footnotesize Electronic states used as basis for discretized Feshbach projection formalism: the localized valence state ($\ket{\Phi_1}$) and a manifold of discretized continuum states ($\ket{\bar{\Phi}_{E_n}}$) coupled to $\ket{\Phi_1}$, where $V_{1E_n}=\bra{\Phi_1}\hat{H}\ket{\bar{\Phi}_{E_n}}$ is the corresponding electronic coupling matrix element.
\protect\label{fig:QD}}
\end{figure}

Identical expressions are obtained by considering a quantum dissipation of a localized state  ($\Phi_1$) into a discretized continuum ($\{\bar{\Phi}_{E_n} \}$, $n\geq2$)~\cite{Nitzanbook}, see Fig.~\ref{fig:QD}. Once the Hamiltonian has the following form:

\begin{equation}
\mathbf{H}=\begin{pmatrix} E_1 & V_{1E_2} & V_{1E_3}  & ... & V_{1E_n} \\  V_{E_21} & E_2 & 0 &  ... & 0 \\ V_{{E_3}1} & 0 & E_3 &  ... & 0 \\ ... & ... & ... & ... & ...\\  V_{{E_n}1} & 0 & 0 & ... &   E_n \end{pmatrix}
\label{eq:Hdiab}
\end{equation}

the resulting expressions for the resonance's energy and width are identical to Eqns.~\ref{eq:width} and ~\ref{eq:energy} provided that $V_{1E_k}=\bra{\Phi_1}H_{QP}\ket{\bar{\Phi}_{E_k}}$ and assuming that the couplings are energy-independent. Therefore, if electronic Hamiltonian is transformed into a basis of a localized resonance and discretized continuum states it is coupled to, the Eqns.~\ref{eq:width} and ~\ref{eq:energy} can be used for evaluating the resonance's parameters. Yet, the electronic states that are eigenstates of electronic Hamiltonian have a mixed localized-discretized continuum character and as such do not form a proper basis for discretized Feshbach projection theory. A procedure  of constructing such as basis from the eigenstates of electronic Hamiltonian would pave the way for using Feshbach projection formalism in conjunction with conventional electronic structure methods for description of electronic structure of autoionizing states. Below we briefly describe extended multiconfigurational  quasidegenerate perturbation theory of second order, the electronic structure method we use in this work, and then proceed to discussion of the proposed procedure of constructing the basis of states needed for Feshbach projection formalism using absorbing potential.
 
 \subsection{XMCQDPT2 model}
 \label{sec:XMCQ}
XMCQDPT2, proposed by A.A. Granovsky~\cite{XMCQDPT2}, is an extension of the original MCQDPT2 model~\cite{MCQDPT2:99:1} model that accounts for invariancy with respect to rotation of the model space vectors. The detailed discussion of the theory and its properties can be found elsewhere~\cite{XMCQDPT2}, here we only summarize the most important practical features relevant for description of electronic resonances. The model space for construction of the effective Hamiltonian up to the second order in perturbation is spanned by several zero-order CASCI states of interest. In case of describing metastable electronic states, the model space is spanned by the states with significant contribution of the localized part of the resonance and the states with the same symmetry lying close in energy. The typical dimensions of the model space in these calculations are 10-20. Diagonalization of the effective Hamiltonian results in perturbation-modified zero-order states, that are linear combination of zero-order states mixed under the influence of dynamic electron correlation. The theory has been previously extended to describing resonance parameters by combining it with CAP approach~\cite{Kunitsa:xmcq:16}. However, the resulting model suffers from the same problems as other CAP-based methods, the need of computing and analyzing $\eta$-trajectories.  The method outlined below allows one to avoid evaluation of $\eta$-trajectories and extracting resonance parameters directly from the effective Hamiltonian matrix.

\subsection{Feshbach-projection-XMCQDPT2 (FP-XMCQDPT2) theory for resonances}
\label{sec:cap-xmcq}
Direct use of Eqns.~\ref{eq:width} and~\ref{eq:energy} for calculation of resonance position and width is only appropriate when a basis of the localized state representing the resonance and a set of discretized continuum states has been obtained and the Hamiltonian has the form as in Eq.~\ref{eq:Hdiab}.  Hereafter we will refer to this basis as projected. As mentioned above, the eigenstates of electronic Hamiltonian are of a mixed localized-discretized continuum character. Below we describe a procedure of generating projected basis from the eigenstates of electronic Hamiltonian using absorbing potential. The method is first illustrated for model one-dimensional potential of the following form:

\begin{equation}
V(x)=\left (1-\frac{1}{\cosh(x^2)}\right )\exp(-0.05x^2)
\label{eq:RMP}
\end{equation} 

\begin{figure}[!tbh]
\includegraphics[width=5cm]{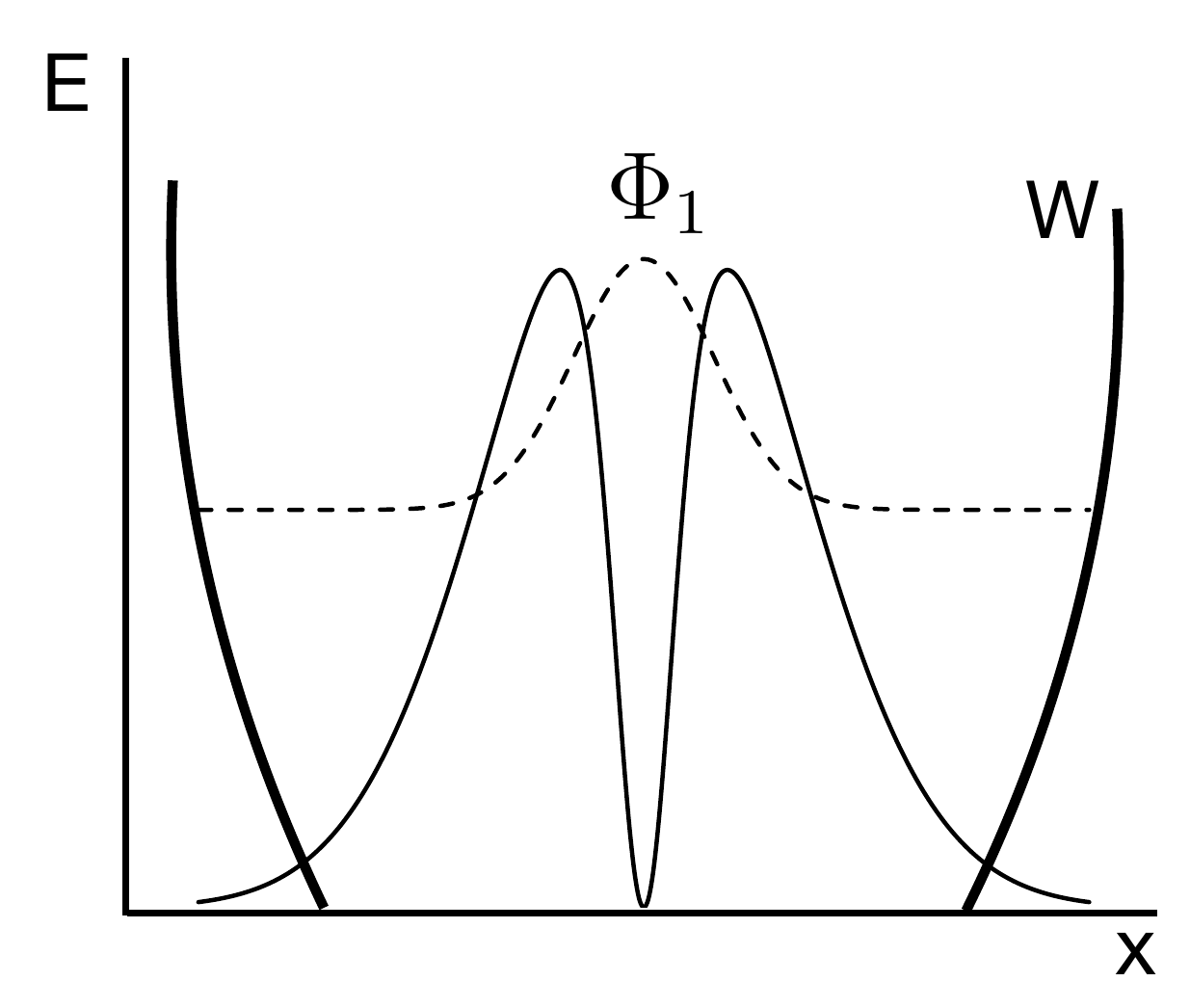}
\caption{\footnotesize Schematic representation of the spatial extent of the localized valence state ($\Psi_d$, dashed), a model potential ($V(x)$, solid) and absorbing potential ($W(x)$, bold solid).
\protect\label{fig:cap_mat_el}}
\end{figure}

\noindent This potential shown schematically in Fig.~\ref{fig:cap_mat_el} supports a resonance with the energy and width of 0.465 and 0.00463 a.u., respectively~\cite{Klaiman:res_rev:12}. The same approach is then applied to calculate resonances in molecular systems. The absorbing potential used for the one-dimensional case is defined as follows:

\begin{equation}
W(x)=\begin{cases} 0, |x|<2.0 \\ (|x|-2.0)^2, |x|\geq2.0  \end{cases}
\label{eq:cap1D}
\end{equation}

The key idea of the method can be inferred from Fig.~\ref{fig:cap_mat_el} illustrating the spatial extent of the localized part of the resonance ($\ket{\Phi_1}$) and absorbing potential $W(x)$ (Eq.~\ref{eq:cap1D}). Provided the projected basis is generated, and $\ket{\Phi_1}$ is the localized state representing the resonance, this state will have the dominant amplitude inside the well of $V(x)$ and decay rapidly outside. $W(x)$ in contrast is non-zero only outside of $V(x)$. Thus, the matrix elements of $W$ between localized part of the resonance and any of the continuum states will be approximately equal to zero as there is almost no overlap between resonances represented in the projected basis,  $\Phi_1$,  and  $W(x)$: 

$$
\mathbf{W}_{1E_n}=\braket{\Phi_1|W|\bar{\Phi}_{E_n}}\approx0
$$

\begin{figure}
\includegraphics[width=8.5cm]{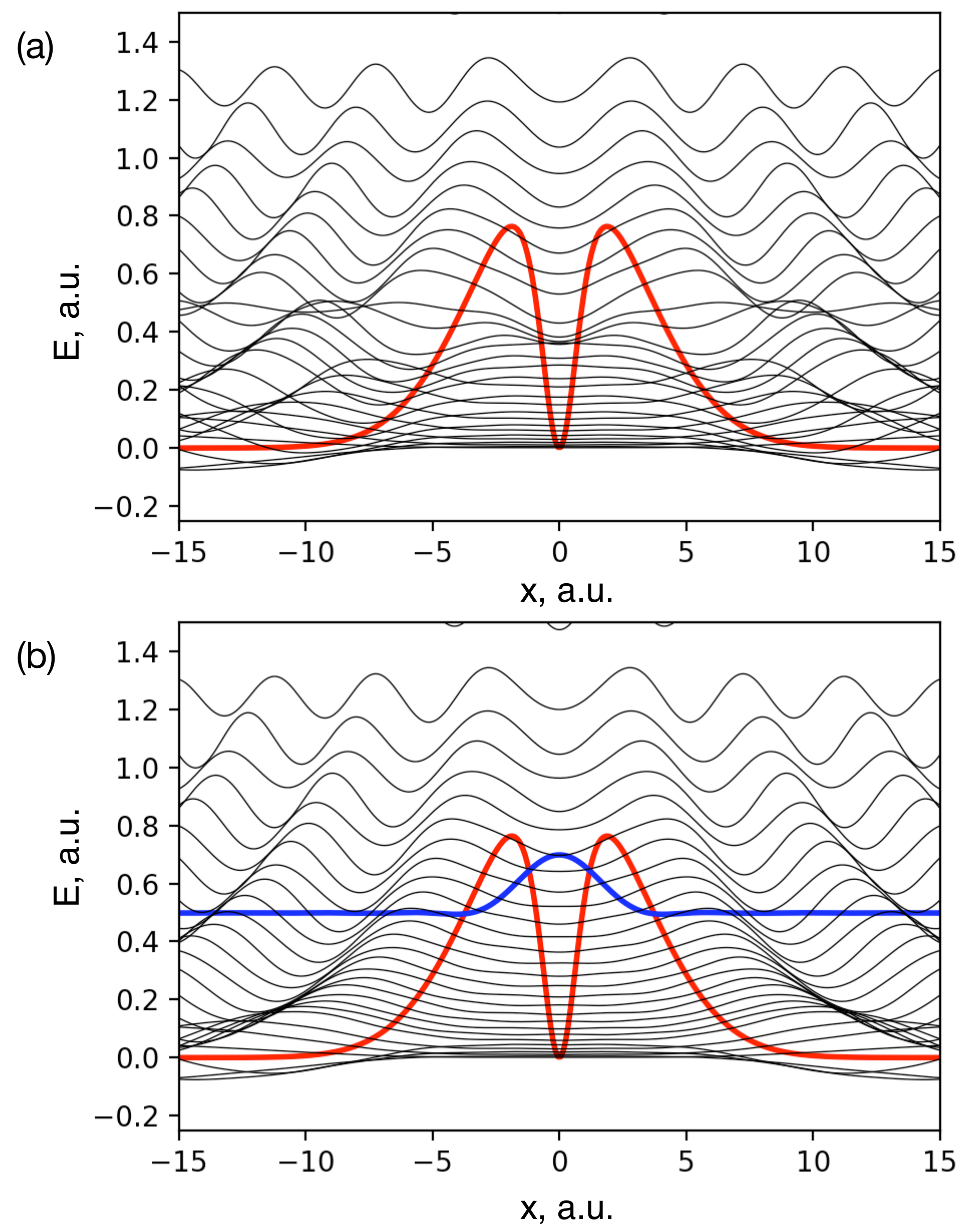}
\caption{\footnotesize Transformation from a mixed localized-continuum basis to projected basis for the 1D model potential (Eq.~\ref{eq:RMP}). The potential is shown in red. (a) Mixed basis of localized and discretized continuum states. Eigenstates of the Hamiltonian are shown in black. (b) Projected basis. Eigenstates of W are shown in black with the localized state representing the resonance ($\ket{\Phi_1}$) shown in blue. 
\protect\label{fig:adaib_diab} }
\end{figure}

Therefore, $\ket{\Phi_1}$ will be an eigenstate of $W$ with almost zero eigenvalue (in practice - minimal eigenvalue). This observation allows one to generate approximated projected basis by diagonalizing $\mathbf{W}$ in the basis of eigenstates of Hamiltonian.  The results of using this procedure for a model 1D potential (Eq.~\ref{eq:RMP}) are shown in Fig.~\ref{fig:adaib_diab}.

A Harmonic oscillator eigenstates basis was used for 1D calculations (30 basis functions, $\omega$=0.025). The resulting eigenstates of Hamiltonian are shown in Fig.~\ref{fig:adaib_diab}a. One can see that there are several states with notable amplitude inside the well. Diagonalization of $\mathbf{W}$ in the basis of eigenstates of the Hamiltonian yields projected basis with single state with a notable amplitude inside the well, $\ket{\Phi_1}$ state, and a manifold of the discretized continuum states, $\{\ket{\bar{\Phi}_{E_n}}\}$, (Fig.~\ref{fig:adaib_diab}b). The Hamiltonian is then transformed into the projected basis, and the continuum states are rotated in such a way that the corresponding block of the Hamiltonian is diagonal and overall Hamiltonian has the form consistent with Eq.~\ref{eq:Hdiab}. The resonance position and width are then evaluated using Eq.~\ref{eq:energy} and~\ref{eq:width}. The estimation of the resonance width requires the coupling matrix element with the continuum at the energy of the localized state. Moreover, the value of density of states at this energy is needed. Here we used the the coupling matrix element for the state that is closest in energy to the localized state and the average value of density of states over the manifold of the discretized continuum states. The resulting estimates of the resonance energy and width are  0.476  and  0.00613 a.u., that are close to the exact numerical results of 0.465 and 0.00463 a.u. Therefore, Feshbach projection formalism together with the outlined procedure of generating projected basis give a good estimate of the resonance position and width for a model 1D problem.

\begin{figure}
\includegraphics[width=8.5cm]{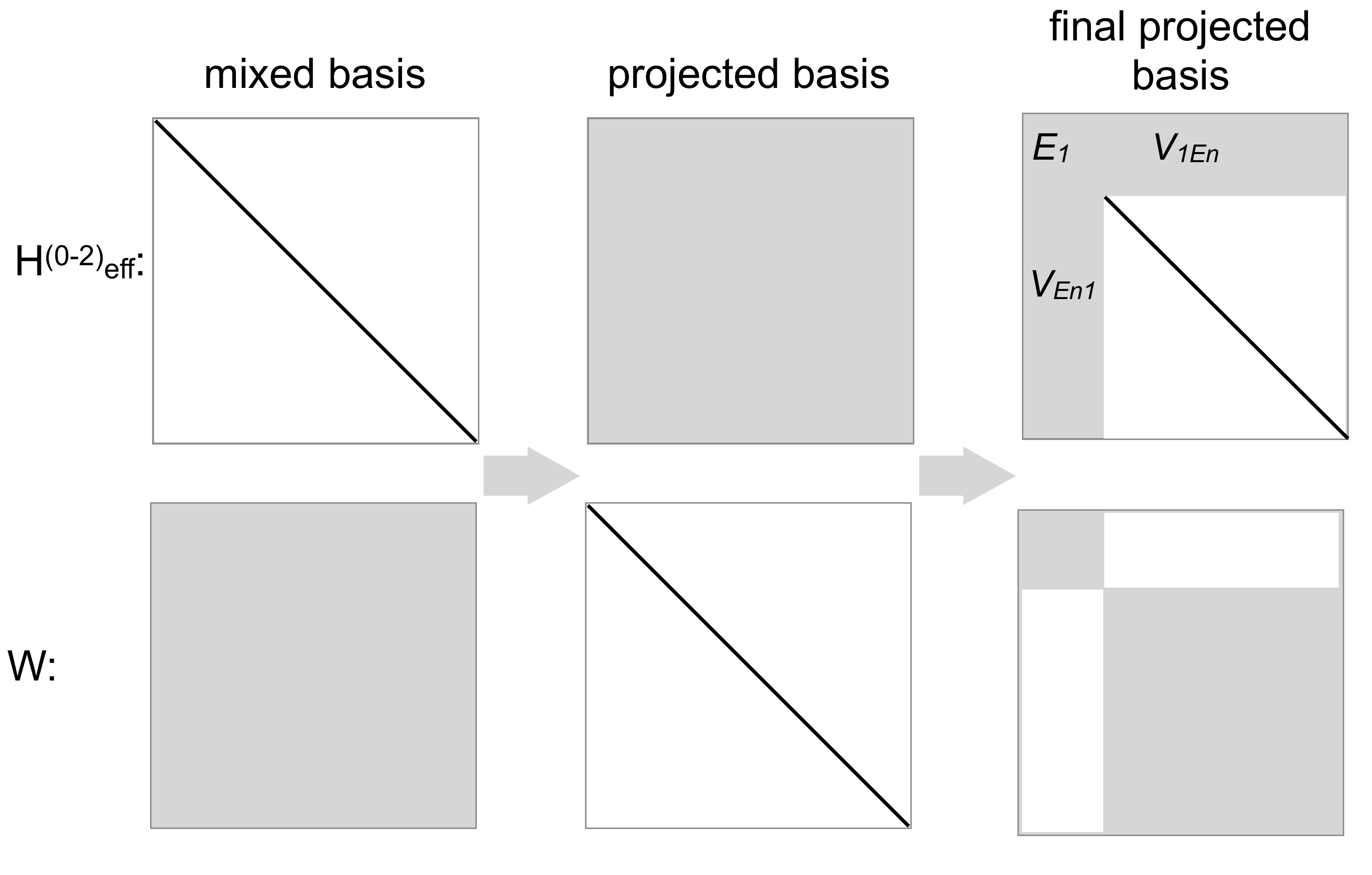}
\caption{\footnotesize Scheme of FP-XMCQDPT2 calculation: $H^{(0-2)}_{eff}$ and $W$ matrices transformation upon generation of the projected basis. 
\protect\label{fig:matrices}}
\end{figure}

The same approach can be generalized to treatment of resonances in molecular systems, for example, by combining with XMCQDPT2 electronic structure method. Note, however, that the approach is transferrable to other quantum chemistry methods.  The scheme of FP-XMCQDPT2 calculation is summarized in Fig.~\ref{fig:matrices}. The first step is a conventional XMCQDPT2 calculation which generates the basis of eigenstates of the effective Hamiltonian, perturbation-modified CASCI states. This basis is a `mixed' basis with the states being of mixed localized -- discretized continuum character. The second step is evaluation in diagonalization of the $W$ matrix in the basis of the the eigenstates of the effective Hamiltonian. This step generates `projected basis', localized state and a manifold of discretized continuum states. Finally, the subset of discretized continuum states should be transformed in such a way that the Hamiltonian matrix is diagonal in this subspace. The resulting Hamiltonian can be used to evaluate resonance position and width using the Eqns. ~\ref{eq:width} and~\ref{eq:energy}.
The approach has been tested on a series of shape resonances in molecular systems as discussed below. The idea of using Fano projection in the context of practical electronic structure calculations for polyatomic molecules has been explored previously, for example for Feshbach resonances, where the projected basis can specified by choosing the states that are generated from a closed-shell reference by a particular class of excitations~\cite{Cederbaum:FanoS:05}. However, this approach is not easily transferable for shape resonances. The scheme proposed here provides a general approach for generating projected basis. While the performance of the model has only been tested for shape resonance, there are no fundamental obstacles that can prevent it's use for Feshbach resonances.

\section{Results and discussion}
\label{sec:results}
To explore the accuracy and stability of the method with respect to parameters of the absorbing potential and basis set the method first applied to well-studied shape resonance in $N_2^-$.   The values of the resonance position and width derived from the experimental data are 2.32 and 0.41 eV, respectively~\cite{Berman:Nitrogen:83}.  Geometries of molecular systems discussed below and details of the basis and model space choice are given in Supporting Information (Secs. S1, S5, and S6) . Additive quadratic-like $W$ was used:

\begin{equation}
W=W_x+W_y+W_z
\nonumber
\end{equation}
\begin{equation}
\label{eq:W}
W_i = \begin{cases} 0, |r_i|\leq R_i \\ (|r_i|-R_i)^2,  |r_i|\geq R_i \end{cases} \\
i=x,y,z,
\end{equation}

The results obtained with traditional CAP-XMCQDPT2 method and FP-XMCQDPT2 for different one electron basis sets with varied number of diffuse basis functions are reported in Table~\ref{table:basis}.

\begin{table}[!tbh]
\caption{N$_2^-$ $^2\Pi_g$ resonance position and width: the effects of the basis set. The following absorbing potential onsets were used: $R_x$= 2.76 bohr,  $R_y$= 2.76 bohr, and $R_z$= 4.88 bohr (Eq.~\ref{eq:W}). 
\label{table:basis}}
\begin{center}
\begin{tabular}{lccccc}
\hline\hline
Method & \multicolumn{2}{c}{CAP-XMCQDPT2 } & \multicolumn{3}{c}{FP-XMCQDPT2} \\
  & $E_R$, eV & $\Gamma$, eV &  $E_R$, eV &  $E_R+\Delta$, eV & $\Gamma$, eV \\
\hline
SA-16-CASSCF(5/18)/ & 1.860 & 0.227 &  2.248 & 1.889 & 0.135 \\
aug-cc-pVTZ+[7s7p7d] & & & & & \\
SA-12-CASSCF(5/14)/ & 2.413 & 0.347 &  2.450 & 2.532 & 0.295 \\
 aug-cc-pVTZ+[5s5p5d] & & & & & \\
SA-8-CASSCF(5/12)/ &  2.512 & 0.385 &  2.621 & 2.572 & 0.215 \\
 aug-cc-pVTZ+[3s3p3d] & & & & & \\
\hline\hline
\end{tabular}
\end{center}
\end{table}

One can see that CAP-XMCQDPT2 and FP-XMCQDPT2 estimates of resonance position and width are in close agreement. For all three bases, the resonance positions for the two methods are within 0.15 eV, and the resonance widths are within 0.2 eV.  Note, however, that one should not expect the values to coincide: CAP-XMCQDPT2 estimates are obtained from eigenstates of CAP-augmented Hamiltonian, a Hamiltonian perturbed by unphysical CAP, whereas in FP-XMCQDPT2 absorbing potential is only used to generate the projected basis of states.

To explore the stability of FP-XMCQDPT2 model to the parameters of absorbing potential we focused on the smaller aug-cc-pVTZ+[3s3p3d] (3 s-, 3 p-, and 3 d- type basis functions added at each nitrogen atom) and varied the absorbing potential onsets from with the increment of 3 bohr (Table~\ref{table:cap}). Both methods exhibit dependence on $W$ onset, however, the resonance position obtained with FP-XMCQDPT2 theory is more stable and varies only from 2.596 to 2.640 eV. In contrast, the resonance position obtained with CAP-XMCQDPT2 method changes more dramatically: from 2.626 to 2.286 eV. Moreover, the $\eta$-trajectory in  CAP-XMCQDPT2 for the largest absorbing potential  does not exhibit a stationary point, and, therefore, the resonance parameters cannot be identified. Interestingly, the FP-XMCQDPT2 model allows one to extract an estimate of the resonance position in this case too, the predicted $E_R$ is 2.640 eV. We also report the computed resonance parameters for $^2\Pi$ resonance in CO$^-$ for different box sizes (see Supporting Information, Sec. S4).

\begin{table}[!tbh]
\caption{N$_2^-$ $^2\Pi_g$ resonance position and width: the effects of variation in the absorbing potential onset. aug-cc-pVTZ+[3s3p3d] basis has been used in all calculations. The model space and active space were the same in all calculations (SA-7-CASSCF(5/13)). 
\label{table:cap}}
\begin{center}
\begin{tabular}{ccccccc}
\hline\hline
$R_x$=$R_y$, bohr &  $R_z$, bohr & \multicolumn{2}{c}{CAP-XMCQDPT2 } & \multicolumn{3}{c}{FP-XMCQDPT2} \\
   & & $E_R$, eV & $\Gamma$, eV &  $E_R$, eV &  $E_R+\Delta$, eV & $\Gamma$, eV \\
\hline
 2.76 & 4.88 & 2.626  &  0.379 &  2.649 &  2.596 &  0.219 \\
5.76 & 7.88 &  2.409  &  0.351 & 2.562  & 2.609   &  0.293 \\
8.76 & 10.88 &  2.286  &   0.203 & 2.500  &2.621    &  0.370 \\
11.76 & 13.88 & --   &  --  & 2.435   &  2.640  &  0.474  \\
\hline
\end{tabular}
\end{center}
\end{table}

The performance of the method was further tested by considering shape resonances in CO$^-$ and polyatomic systems. The results are given in Table~\ref{table:poly}. One can see that in most cases, there is a good agreement between the two models. The first considered shape resonance was $^2\Pi$ resonance of CO$^-$. The experimental estimate of the resonance energy and width for this state are 1.50 and 0.40-0.80 eV, respectively~\cite{Erhardt:Carbonmonoxide:68,Zubek:Carbonmonoxide:77,Zubek:CO:79}.  FP-XMCQDPT2 values for the resonance parameters are in close agreement with the experiment. The consistency between CAP-XMCQDPT2 and FP-XMCQDPT2 models is also observed for resonances in larger molecular systems, such as formic acid, formaldehyde, and methyl formate anions.

 \begin{table}[!tbh]
\caption{Resonance position and width: CO$^-$ and polyatomic anions. 
\label{table:poly}}
\begin{center}
\begin{tabular}{llccccc}
\hline\hline
Resonance & Method & \multicolumn{2}{c}{CAP-XMCQDPT2 } & \multicolumn{3}{c}{FP-XMCQDPT2} \\
& & $E_R$, eV & $\Gamma$, eV &  $E_R$, eV &  $E_R+\Delta$, eV & $\Gamma$, eV \\
\hline
$^2\Pi$, CO$^-$ & SA-9-CASSSCF(3/10)/  & 1.547  & 0.479 & 1.980 &   1.811 &    0.411 \\
& aug-cc-pV5Z+[1gh:3s3p3d] & & & & & \\
$^2A''$ ($\pi^*$), formic acid anion & SA-12-CASSCF(3/13)/&  1.950 & 0.193 & 2.350 & 1.938 &  0.237 \\
& cc-pVTZ+[6gh:3s] & & & & & \\
$^2B_2$ ($\pi^*$), formaldehyde anion & SA-10-CASSCF(3/9)/& 1.080 & 0.155 & 1.545 & 1.232 & 0.038 \\
  & cc-pVTZ+[4gh:3s] & & & & & \\
$^2A''$ ($\pi^*$), methylformate anion & SA-12-CASSCF(3/11)/& 2.340 & 0.189 & 2.495 & 2.461 & 0.152 \\
& cc-pVTZ+[6gh:3s] & & & & & \\
\hline
\end{tabular}
\end{center}
\end{table}

While the model yields accurate results for resonance position and width for selected shape resonances in molecules, some fo the features of the method should be further explored. Specifically, one can expect problematic behavior of the technique when very diffuse bases are used (and therefore large model space spanned by many closely-spaced discretized continuum states). Owing to perturbative nature of the resonance shift expression (Eq.~\ref{eq:energy}) one can anticipate divergent behavior for the resonance energy. Careful assessment of the basis set effects, analysis of continuity of the complex potential energy surfaces, and performance for the Feshbach resonances are the subject of the future work.

\section{Conclusions}
We present a new model that combines Feshbach projection formalism with XMCQDPT2 model to enable treatment of resonance electronic states. The method yields resonance position and width at a cost of single electronic structure calculation and does not require evaluation and analysis of $\eta$-trajectories. In most of the considered cases, the FP-XMCQDPT2 model yields estimates of resonance position and width that are in good agreement with the corresponding CAP-XMCQDPT2 values. FP-XMCQDPT2 performance is expected to be problematic when very diffuse basis are involved owing to quasidegeneracy between the localized state and one or several continuum states. However, the model has been shown to yield a reliable estimates of resonance position and width in some cases, where conventional CAP-XMCQDPT2 model fails. \label{sec:conclusions}
 
 \section*{Acknowledgements}
 This research was supported by Army Research Office (W911NF1910072).


\clearpage
%
%

\draft 
\renewcommand{\thefigure}{S\arabic{figure}}
\renewcommand{\thetable}{S\arabic{table}}
\renewcommand{\thesection}{S\arabic{section}}
\renewcommand{\theequation}{S\arabic{equation}}
 
\renewcommand{\thepage}{S\arabic{page}}
\setcounter{section}{0}


\begin{center}
{\it \large Supporting Information for: 
Feshbach projection XMCQDPT2 model for metastable electronic states}



Alexander A. Kunitsa and Ksenia B. Bravaya \\
Department of Chemistry, Boston University, Boston, Massachusetts, \nolinebreak{02215, USA}
\end{center} 


\section{Active space choice}
The initial active space selection for each CASSCF calculation was based on RHF orbitals of a neutral molecule. The orbitals of $\pi^*$ character, all orbitals of the same symmetry lying energetically below, and several orbitals of the same symmetry lying energetically above were included into the active space. The active space also included one very diffuse `fake-ip' orbital (exponent of 1$\times 10^{-12}$) to mimic ionization. The latter allows one to treat the states with N+1 and N electrons within the same state-averaged calculation (the lowest state in each calculation is the ground state of the neutral (N-electron system) with the extra electron at the `fake-ip' orbital.  
\section{Resonance shift evaluation: quasidegeneracy}
To avoid divergencies due to quasidegenracy of the localized part of the resonance and discretized continuum states arising from the perturbative expression for the resonance shift as shown in the following expression,

$$
E_{R}=E_1+\sum_{k\neq1}{\frac{|V_{1E_k}|^2}{E_1-E_k}}
$$
the contribution of the states that are within 0.2 eV from the energy of the localized state ($E_1$) were neglected in the sum above.
\section{Absorbing potential onsets}
Unless stated otherwise the onsets of the absorbing potentials listed in Table~\ref{table:onsets} were used in the calculations.
\begin{table}[!tbh]
\caption{Absorbing potenital onsets
\protect\label{table:onsets}}
\begin{tabular}{lccc}
\hline \hline
Molecule and state & R$_x$, bohr & R$_y$, bohr & R$_z$, bohr \\
\hline
N$_2^-$ $^2\Pi_g$ &  2.76 & 2.76 & 4.88 \\
CO $^2\Pi$ & 2.76 & 2.76 & 4.97 \\
formic acid $^2A"$ & 8.00 & 8.00 & 8.00 \\
formaldehyde $^2B_2$ & 8.00 & 8.00 & 8.00 \\
methyl formate $^2A"$ & 8.00 & 8.00 & 8.00 \\
\hline
\end{tabular}
\end{table}

\section{Effects of the box size on resonance position and width: CO$^-$ $^2\Pi$ resonance}
As follows from Tables~\ref{table:cap} and ~\ref{table:cap1}, the FP-XMCQDPT2 yields resonance positions and widths that are close to those obtained with CAP-XMCQDPT2 method for more compact absorbing potentials. For wider absorbing potential, the projection approach does not reliably separate the localized state from the discretized continuum state and the procedure fails. Moreover, one can see that the procedure is more stable for less diffuse one-electron basis (Table~\ref{table:cap}), which is consistent with the discussion in the main text.

\begin{table}[!tbh]
\caption{CO$^-$ $^2\Pi$ resonance position and width: the effects of variation in the absorbing potential onset. aug-cc-pV5Z+[4d] basis has been used in all calculations. The [4d] diffuse basis functions were located at the ghost atom in the center of mass.  The model space and active space were the same in all calculations (SA-10-CASSCF(3/11)). Three states out of 10 were treated as localized.
\label{table:cap}}
\begin{center}
\begin{tabular}{lccccc}
\hline\hline
$R_x=R_y=R_z$ onset, bohr & \multicolumn{2}{c}{CAP-XMCQDPT2 } & \multicolumn{3}{c}{FP-XMCQDPT2} \\
  & $E_R$, eV & $\Gamma$, eV &  $E_R$, eV &  $E_R+\Delta$, eV & $\Gamma$, eV \\
\hline
 2.0 & 1.55 &   0.552 &  1.99  &  1.49 &   0.575 \\
4.0 & 1.49  & 0.394 &   1.94  &  1.56   &  0.552 \\
6.0 & 1.43 &   0.253 &     1.92  &  1.59 &    0.551 \\
8.0 & 1.39  &  0.137 & --  &  --  &   -- \\
10.0 & 1.35 &  0.083 & --  &  --   & --  \\
\hline
\end{tabular}
\end{center}
\end{table}

\begin{table}[!tbh]
\caption{CO$^-$ $^2\Pi$ resonance position and width: the effects of variation in the absorbing potential onset. aug-cc-pV5Z+[3d] basis has been used in all calculations. The [3d] diffuse basis functions were located at the ghost atom in the center of mass.  The model space and active space were the same in all calculations (SA-8-CASSCF(3/9)). Two states out of 8 were treated as localized.
\label{table:cap1}}
\begin{center}
\begin{tabular}{lccccc}
\hline\hline
$R_x=R_y=R_z$ onset, bohr & \multicolumn{2}{c}{CAP-XMCQDPT2 } & \multicolumn{3}{c}{FP-XMCQDPT2} \\
  & $E_R$, eV & $\Gamma$, eV &  $E_R$, eV &  $E_R+\Delta$, eV & $\Gamma$, eV \\
\hline
2.0 & 1.60  &  0.552 & 2.00 & 1.81 & 0.402 \\
4.0 & 1.54 &   0.399 & 1.95 &    1.82 &    0.399 \\
6.0 &  1.47 &    0.260 &  1.91  &  1.83 &   0.401 \\
8.0 &  1.42 &   0.142 & 1.94 &    1.82 &     0.429 \\
10.0 &1.39  &   0.083 & -- & -- & --\\
\hline
\end{tabular}
\end{center}
\end{table}

\clearpage
\section{Equilibrium geometries}
Equilibrium geometries of all model systems, except N$_2$, used in the work were optimized with RIMP2/aug-cc-pVTZ. N$_2$ geometry was taken from Ref. 21 (bond length of  1.098 \AA). The atomic coordinates (in \AA) for each of the molecules used in this work are listed below (\AA).

\noindent {\em N$_2$ equilibrium geometry}
\begin{verbatim}
N 0.000000 0.000000 0.548757
N 0.000000 0.000000 -0.548757
\end{verbatim}
\noindent {\em CO equilibrium geometry}
\begin{verbatim}
  C         0.0000000000    0.0000000000   -0.6599743590
  O         0.0000000000    0.0000000000    0.4742600733
\end{verbatim}
\noindent {\em Formic acid equilibrium geometry}
\begin{verbatim}
  C         0.1534580779   -0.3831210567    0.0000000000
  H         0.2096338166   -1.4698711227    0.0000000000
  O         1.0973008256    0.3596222935    0.0000000000
  O        -1.1324942501   -0.0006829175    0.0000000000
  H        -1.1343281591    0.9675708786    0.0000000000
\end{verbatim}
\noindent {\em Formaldehyde equilibrium geometry}
\begin{verbatim}
  C         0.5488245108   -0.0043302130    0.0000000000
  O        -0.6569894499   -0.0875674787    0.0000000000
  H         1.1885948956   -0.8945743045    0.0000000000
  H         1.0603366149    0.9657492628    0.0000000000
\end{verbatim}
\noindent {\em Methyl formate equilibrium geometry}
\begin{verbatim}
  O         0.4816050635   -0.7359419360    0.0000000000
  C        -0.8196786869   -0.4375975282    0.0000000000
  O        -1.2857561038    0.6723968206    0.0000000000
  H        -1.3984996877   -1.3605338374    0.0000000000
  C         1.3444030595    0.4094961496    0.0000000000
  H         2.3514526092    0.0152913875    0.0000000000
  H         1.1657575693    1.0113177988    0.8841979311
  H         1.1657575693    1.0113177988   -0.8841979311
\end{verbatim}

\section{Basis sets}
The {\em diffuse subsets} along with the coordinates of ghost atoms (if any) used in the calculations are listed below.

 {\bf \em N$_2^-$ $^2\Pi_g$ resonance.}

One ghost atom with a corresponding (3s3p3d, 5s5p5d, or 7s7p7d) subset of diffuse basis functions was placed at the center of mass. 'Fake-ip' d-type diffuse orbital with the exponent of 1$\times$10$^-12$ was also centered at the ghost atom.

\noindent {3s3p3d+fake-ip}
\begin{verbatim}
   S          1
     1         0.0288000000  1.00000000
   S          1
     1         0.019200000  1.00000000
   S          1
     1         0.012800000  1.00000000
   P          1
     1         0.0245500000  1.00000000
   P          1
     1         0.0163666666  1.00000000
   P          1
     1         0.0081833333  1.00000000
   D          1
     1         0.0755000000  1.00000000
   D          1
     1         0.0503333333  1.00000000
   D          1
     1         0.0335555555  1.00000000
   D          1
     1         0.000000000001  1.00000000
\end{verbatim}

\noindent {5s5p5d+fake-ip}
\begin{verbatim}
   S          1
     1         0.0288000000  1.00000000
   S          1
     1         0.019200000  1.00000000
   S          1
     1         0.012800000  1.00000000
   S          1
     1         0.008533333  1.00000000
   S          1
     1         0.005688888  1.00000000
   P          1
     1         0.0245500000  1.00000000
   P          1
     1         0.0163666666  1.00000000
   P          1
     1         0.0081833333  1.00000000
   P          1
     1         0.0061375000  1.00000000
   P          1
     1         0.0040966666  1.00000000
   D          1
     1         0.0755000000  1.00000000
   D          1
     1         0.0503333333  1.00000000
   D          1
     1         0.0335555555  1.00000000
   D          1
     1         0.02237037037  1.00000000
   D          1
     1         0.01491358025  1.00000000
   D          1
     1         0.000000000001  1.00000000
\end{verbatim}

\noindent {7s7p7d+fake-ip}
\begin{verbatim}
  S          1
     1         0.0288000000  1.00000000
   S          1
     1         0.019200000  1.00000000
   S          1
     1         0.012800000  1.00000000
   S          1
     1         0.008533333  1.00000000
   S          1
     1         0.005688888  1.00000000
   S          1
     1         0.002844444  1.00000000
   S          1
     1         0.001444444  1.00000000
   P          1
     1         0.0245500000  1.00000000
   P          1
     1         0.0163666666  1.00000000
   P          1
     1         0.0081833333  1.00000000
   P          1
     1         0.0061375000  1.00000000
   P          1
     1         0.0040966666  1.00000000
   P          1
     1         0.0020483333  1.00000000
   P          1
     1         0.0010241667  1.00000000
   D          1
     1         0.0755000000  1.00000000
   D          1
     1         0.0503333333  1.00000000
   D          1
     1         0.0335555555  1.00000000
   D          1
     1         0.02237037037  1.00000000
   D          1
     1         0.01491358025  1.00000000
   D          1
     1         0.00745679012  1.00000000
   D          1
     1         0.00372839506  1.00000000
   D          1
     1         0.000000000001  1.00000000
\end{verbatim}

 {\bf \em CO$^-$ $^2\Pi$ resonance.}

One ghost atom with a corresponding (3d or 4d) subset of diffuse basis functions was placed at the center of mass. 'Fake-ip' d-type diffuse orbital with the exponent of 1$\times$10$^-12$ was also centered at the ghost atom.

\noindent {3d+fake-ip}
\begin{verbatim}
   D          1
     1         0.0350500000  1.00000000
   D          1
     1         0.0175250000  1.00000000
   D          1
     1         0.0087625000  1.00000000
   P          1
     1         0.000000000001  1.00000000
\end{verbatim}

\noindent {4d+fake-ip}
\begin{verbatim}
   D          1
     1         0.0350500000  1.00000000
   D          1
     1         0.0175250000  1.00000000
   D          1
     1         0.0087625000  1.00000000
   D          1
     1         0.0043812500  1.00000000
   P          1
     1         0.000000000001  1.00000000
\end{verbatim}

 {\bf \em $^2A"$ resonances in formic acid anion.}
 
 Six ghost ghost atoms were added above and below molecular plane at positions of heavy atoms. The coordinates of the ghost atoms are listed below:
 
\begin{verbatim}
Gh  0.1534580779     -0.3831210567      1.0000000000
Gh  0.1534580779     -0.3831210567     -1.0000000000
Gh  1.0973008256      0.3596222935      1.0000000000
Gh  1.0973008256      0.3596222935     -1.0000000000
Gh -1.1324942501     -0.0006829175      1.0000000000
Gh -1.1324942501     -0.0006829175     -1.0000000000
\end{verbatim}
 
 Each ghost atom hosted 3 s-type diffuse basis functions:
\begin{verbatim}
   S          1
     1         0.0220100000  1.00000000
   S          1
     1         0.0110050000  1.00000000
   S          1
     1         0.0055025000  1.00000000
\end{verbatim}
 
  'Fake-ip' d-type diffuse orbital with the exponent of 1$\times$10$^-12$ was also centered at the ghost atom.
  
   {\bf \em $^2B_2$ resonances in formaldehyde anion.}
 
 Four ghost ghost atoms were added above and below molecular plane at positions of heavy atoms. The coordinates of the ghost atoms are listed below:
 
\begin{verbatim}
Gh      0.5488245108     -0.0043302130      1.0000000000
Gh      0.5488245108     -0.0043302130     -1.0000000000
Gh     -0.6569894499     -0.0875674787      1.0000000000
Gh     -0.6569894499     -0.0875674787     -1.0000000000
\end{verbatim}
 
 Each ghost atom hosted 3 s-type diffuse basis functions:
\begin{verbatim}
   S          1
     1         0.0220100000  1.00000000
   S          1
     1         0.0110050000  1.00000000
   S          1
     1         0.0055025000  1.00000000
\end{verbatim}
 
  'Fake-ip' d-type diffuse orbital with the exponent of 1$\times$10$^-12$ was also centered at the ghost atom.
  
     {\bf \em $^2A_"$ resonances in methyl formate.}
 
 Six ghost ghost atoms were added above and below molecular plane at positions of three heavy atoms (two oxygens and carbonyl carbon). The coordinates of the ghost atoms are listed below:
 
\begin{verbatim}
Gh     0.4816050635     -0.7359419360      1.0000000000
Gh     0.4816050635     -0.7359419360     -1.0000000000
Gh    -0.8196786869     -0.4375975282      1.0000000000
Gh    -0.8196786869     -0.4375975282     -1.0000000000
Gh    -1.2857561038      0.6723968206      1.0000000000
Gh    -1.2857561038      0.6723968206     -1.0000000000
\end{verbatim}
 
 Each ghost atom hosted 3 s-type diffuse basis functions:
\begin{verbatim}
   S          1
     1         0.0220100000  1.00000000
   S          1
     1         0.0110050000  1.00000000
   S          1
     1         0.0055025000  1.00000000
\end{verbatim}
 
  'Fake-ip' d-type diffuse orbital with the exponent of 1$\times$10$^-12$ was also centered at the ghost atom.


%

\end{document}